\documentclass{iau}

\usepackage{amsmath}
\usepackage{caption}
\usepackage{subcaption}
\usepackage{graphicx}
\usepackage{hyperref}
\usepackage{multirow}
\begin{document}

\lefttitle{Kelahan et al.}
\righttitle{A Machine Learning Based Search for Lunar Anomalies}

\jnlPage{1}{7}
\jnlDoiYr{2026}
\doival{10.1017/xxxxx}

\aopheadtitle{Proceedings IAU Symposium}
\editors{J. Haqq-Misra \&  R. Kopparapu, eds.}

\title{A Machine Learning Based Search for Lunar Anomalies}

\author {Cameron Kelahan$^{1,2,3}$, Daniel Angerhausen$^4$,\\
         Adam Lesnikowski$^{5}$, and Valentin T. Bickel$^{6}$}

\affiliation{
$^1$University of Warwick, Coventry, United Kingdom CV4 7AL \\
$^2$Universit\`a degli Studi di Padova, Via 8 Febbraio, 2 - 35122 Padova, Italy \\
$^3$James Madison University, 800 South Main Street, Harrisonburg, VA 22807 \\
$^4$The SETI Institute, 339 Bernardo Ave, Suite 200, Mountain View, CA 94043, United States \\
$^5$University of California, Berkeley, 910 Evans Hall, Berkeley, CA 94720 \\
$^6$Center for Space and Habitability, University of Bern, Gesellschaftsstrasse 6, 3012 Bern, Switzerland
}

\begin{abstract}
The Lunar Reconnaissance Orbiter (LRO) has been collecting high-resolution images (at $\sim0.5$-$2$ m/pixel linearly with its Narrow Angle Camera) of the Moon since $2009$, amassing a large dataset of images and offering researchers the opportunity to study the surface of the Moon at unprecedented scale. Here, we aim to test the abilities of the $\beta$-Variational Autoencoder (VAE) created by \cite{lesnikowski2024anomalies}, an unsupervised learning model which identifies anomalous features across the Moon’s surface, locating not only scientifically useful geologic formations such as rockfall deposits, fresh impact craters, irregular mare patches, or volcanic pits/collapsed lava tubes, but also artificial objects such as landed spacecraft. This investigation further gauged the model's ability to locate anomalous surface features, successfully recovering two places of interest (Plaskett Crater and Paracelsus C Crater) and numerous landed technological assets at a statistically significant rate.
\end{abstract}


\maketitle

\vspace{-1em}

\section{Introduction}

The Moon has captured the attention of humanity for its entire existence, holding both cultural and scientific importance, culminating in its hands-on exploration in the last century. From ancient civilizations tracking the Moon and its phases, to early observers utilizing rudimentary telescopes to record lunar features, to modern high-resolution image mapping of the surface, the Moon has been the center of studies for millennia. Learning what the Moon has to share has advanced the understanding of planetary formation, impact processes, the history of Earth, and of our Solar System in general. One of the inspirations for this project came from \cite{davies2013searching} who argued for a manual search of LRO images for anything that may have been left behind by aliens. Another was a desire to re-examine our nearest space-neighbor with modern tools that offer new perspectives, looking to answer the question: could there be technosignatures on the Moon?

The dataset we used in this project contains images taken by the Lunar Reconnaissance Orbiter (LRO) over the last 17 years. The LRO was launched by NASA on June $18$, $2009$ with the goal of producing detailed mapping and remote sensing of the Moon's surface. Looking up at the Moon, even with the naked eye, its diverse surface of large maria, highlands, and craters can be observed. The LRO offers a much closer perspective, allowing for the observation of finer details; this includes the likes of interesting surface anomalies such as irregular mare patches (IMPs), fresh impact craters, rockfalls, volcanic pits/collapsed lava tubes, other geological formations, and even landed technological assets.

In this work we examined the capabilities of the model introduced by \cite{lesnikowski2024anomalies} which highlighted the viability of new science waiting to be discovered by re-examining existing datasets through the lens of cutting edge technology and advancements such as generative machine and deep learning models. In general, there are 2 major types of machine learning models: supervised and unsupervised. Supervised learning models are given both the raw data and the true corresponding classification or label for the data \citep[e.g.][]{Pinault2026}. The model in this project avoids the use of labeled data, which often relies on a laborious, and biased, process to obtain, not by solely comparing pixel color values as done in standard computer vision algorithms \citep[e.g.][]{vision6030054}, but by employing the use of unsupervised learning where the algorithm learns a domain without explicitly being told what it is processing. It does this by uncovering underlying patterns and connections within the training set. Due to the Moon’s overarching homogeneity, any kind of surface anomaly stands out when compared to what the ‘average’ lunar surface looks like, allowing the algorithm to quantitatively rate surface features by comparing their appearance to the mathematically derived `average lunar surface.'

\section{Methodology}

\begin{figure}[!htbp]
\centering
\begin{subfigure}[t]{0.325\linewidth}
    \centering
    \includegraphics[width=\linewidth]{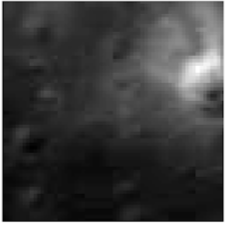}
    \caption{}
    \label{subfig:raw-image-patch}
\end{subfigure}
\begin{subfigure}[t]{0.325\linewidth}
    \centering
    \includegraphics[width=\linewidth]{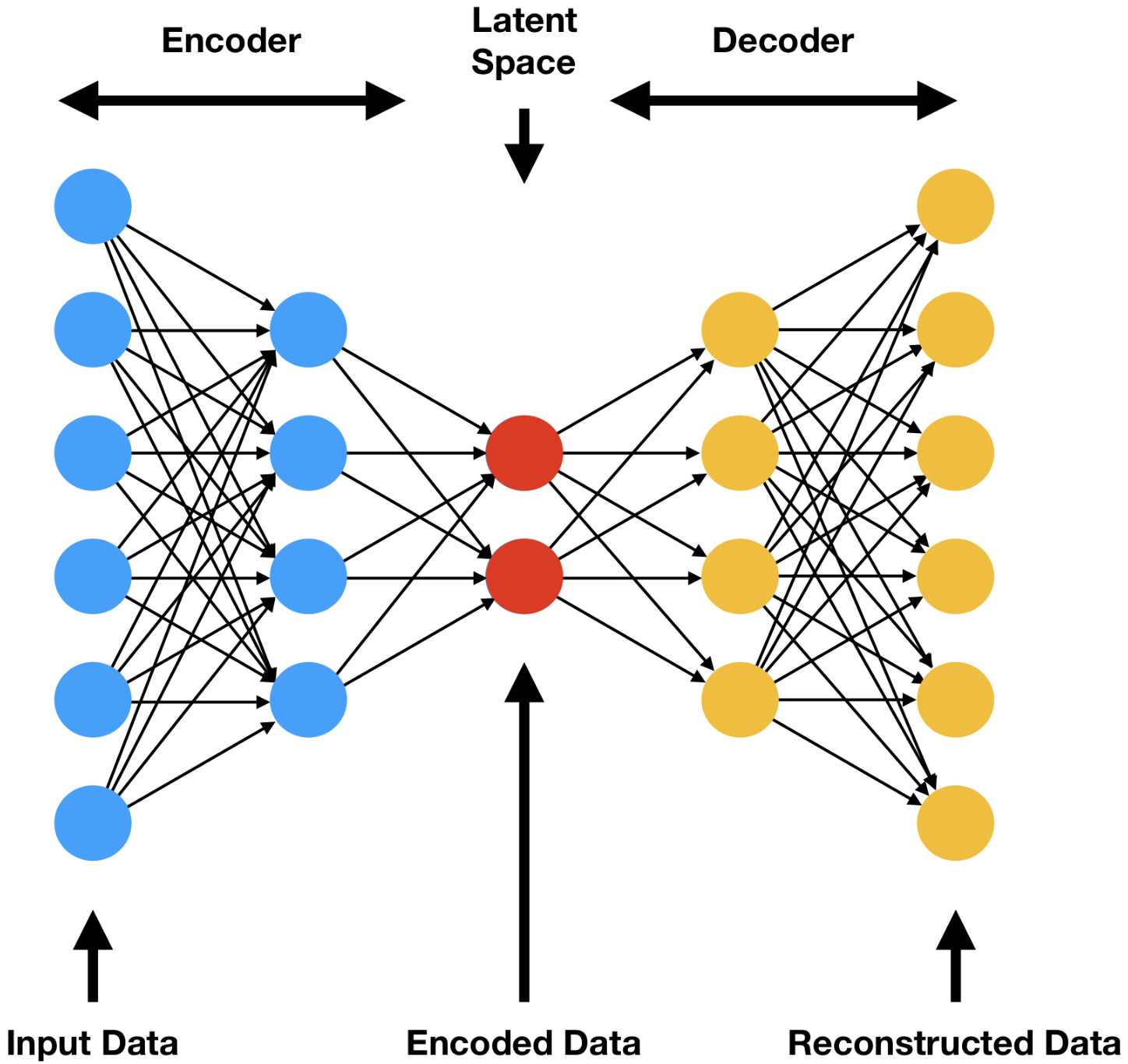}
    \caption{}
    \label{subfig:autoencoder_diagram}
\end{subfigure}
\begin{subfigure}[t]{0.325\linewidth}
    \centering
    \includegraphics[width=\linewidth]{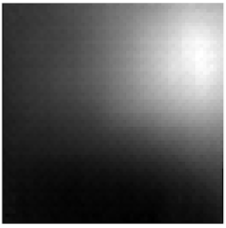}
    \caption{}
    \label{subfig:reconstructed_image}
\end{subfigure}
\caption{An example of a raw patch (64x64 pixels) (a) and its reconstruction (c) with a generic example of an autoencoder architecture between them (b) \citep{seohyun2023autoencoder}}
\label{fig:vae_reconstructed_example}
\end{figure}

We employ a $\beta$-Variational Autoencoder \citep{higgins2017betaVAE} trained on a dataset of LRO images \citep[more detailed information can be found in][]{lesnikowski2024anomalies}, learning the latent (underlying) representation of the statistically normal lunar surface. The LRO images are very large ($\sim$50,000 x 5,000 pixels; each pixel is nominally 0.5m x 0.5m), so the data fed to the model is split into patches of a specified size for computational efficiency. This model's patch size was chosen to be 64x64 pixels ($\sim$32x32 m) to strike a balance between capturing large anomalies (caves, craters, etc.) and small anomalies (landed technological assets). The training dataset consisted of $\sim$1,000 images randomly selected across the Moon, totaling over 52 million patches to train on. The model ingested these patches, learning underlying patterns and connections between them, and then attempted to recreate them one at a time based on what it had learned - what it expects the lunar surface to look like (see \autoref{fig:vae_reconstructed_example}). The model does not create a high-definition, perfect recreation of the input image. This  behavior comes from the `$\beta$' regularization term, which alters the objective function to put more emphasis on capturing the underlying features of the data, rather than focusing on exact reconstructions.

By comparing the reconstructed image patch to the original image patch on a pixel-by-pixel level, the amount in which they differ is quantified as an ``anomaly score." Patches with high anomaly scores exhibit anomalous features, such as volcanic pits/collapsed lava tubes (see \autoref{subfig:pits}), while less anomalous patches exhibit fairly homogeneous regolith. While there is an inherent bias towards anomalies that take up the majority of a patch, the metric still manages to identify smaller anomalies (see \autoref{subfig:apollo_16} and \autoref{subfig:change_6}).

\section{Results}

\begin{figure}
    \centering
    \includegraphics[width=0.48\linewidth]{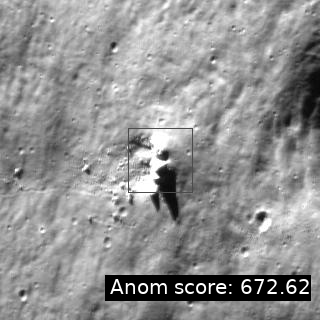}
    \includegraphics[width=0.48\linewidth]{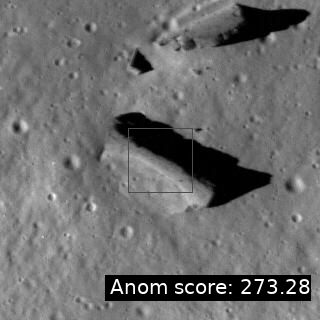}
    \caption{The Plaskett Crater (left) and Paracelsus C (right) anomalies with their respective anomaly scores as determined by the model. As seen in \autoref{subfig:all_tech_anom_histogram}, the average anomaly score of all examined patches was $\sim$23, showing how the model was easily able to detect these anomalous surface features.}
    \label{fig:plaskett_paracelsus}
\end{figure}

To further test the performance of this model, we examined two community-discovered anomalies-of-interest to determine if the model would assign them significant anomaly scores. The chosen sites were Plaskett Crater and Paracelcus C Crater (see \autoref{fig:plaskett_paracelsus}). The model was given LRO images that contained these anomalies and calculated the anomaly scores for each 64x64 pixel patch of the images. The model successfully identified patches containing these two targets of interest as anomalous, displaying its ability to highlight unique surface features, such as these two rock formations. We note that the Plaskett anomaly is a bedrock outcrop on a slope that has been acting as a source of rockfalls (see boulder track leading towards the left); the Paracelsus C anomaly includes two angular boulders that are most likely remnants of a past impact event.

\begin{figure}[!htbp]
  \centering
  \begin{subfigure}[t]{0.49\textwidth}
    \centering
    \includegraphics[width=\textwidth]{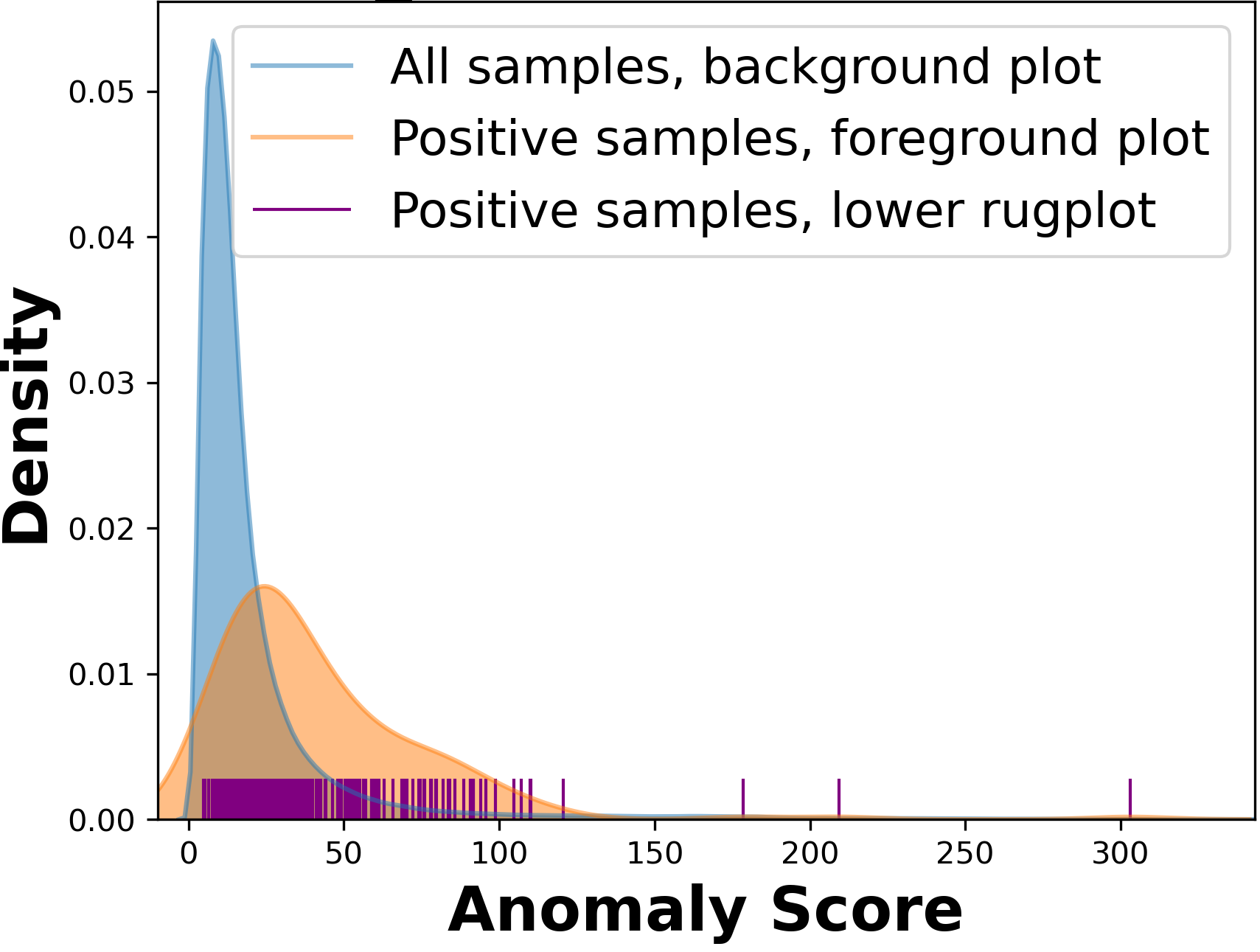}
    \caption{}
    \label{subfig:all_tech_anom_histogram}
  \end{subfigure}
  \hfill
  \begin{subfigure}[t]{0.49\textwidth}
    \includegraphics[width=\textwidth]{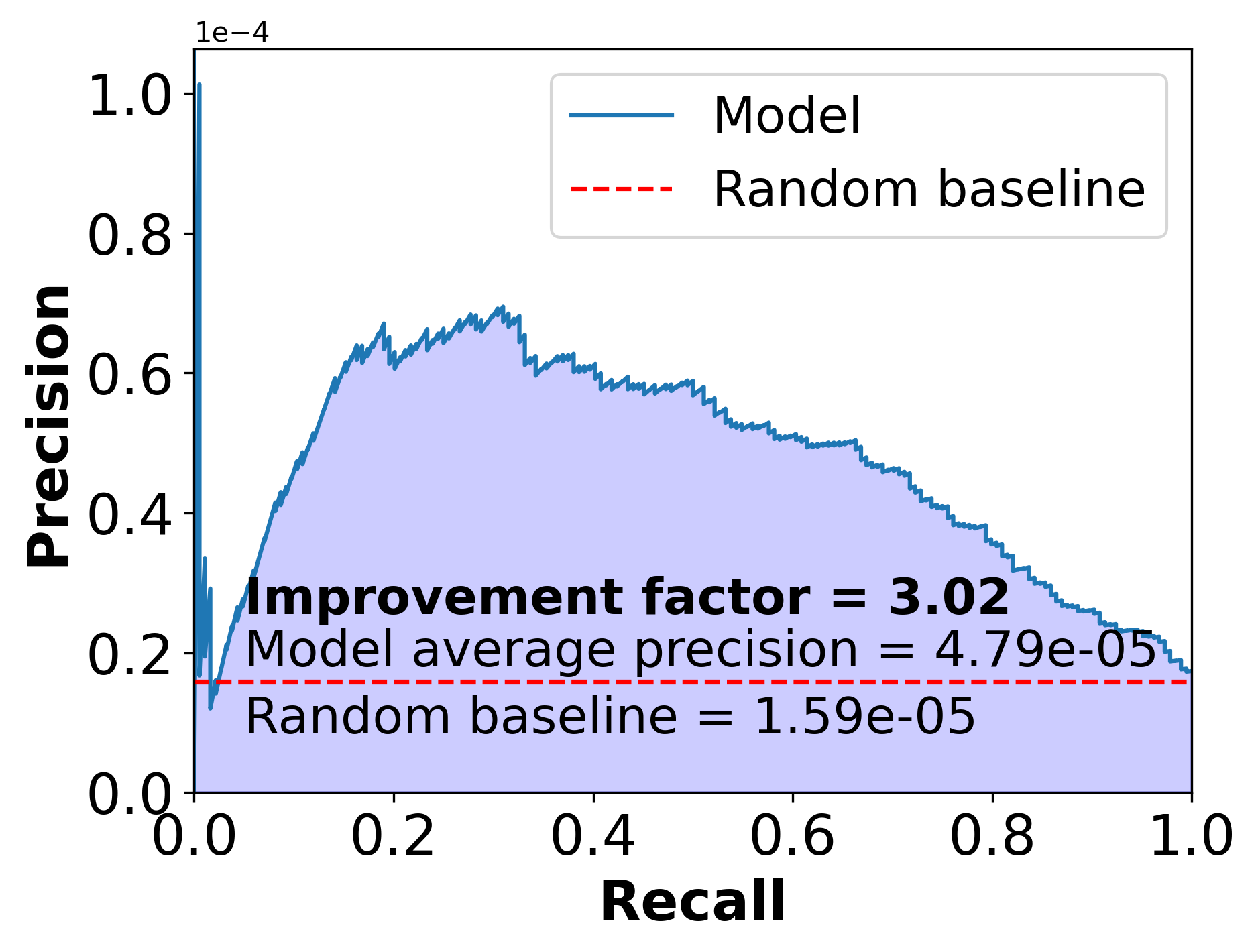}
    \caption{}
    \label{subfig:all_tech_anom_kde}
  \end{subfigure}
  \caption{The anomaly score distribution plot (a) and Precision-Recall curve (b) for all of the images used in this project that captured a technosignature anomaly. The Kolmogorov-Smirnov test \citep{smirnov1948} produced a value of $0.468$ with a p-value of $1.95$x$10^{-35}$, heavily supporting two different distributions of patches with \textit{known} anomalies and patches without \textit{known} anomalies (although they may still contain anomalies) The model's average precision improved by a factor of $3.02$ when compared to the random baseline average precision.}
  \label{fig:all_tech_anom_pr_and_kde}
\end{figure}

\begin{figure}[!hb]
\centering
\begin{subfigure}[t]{0.48\linewidth}
    \centering
    \includegraphics[width=\linewidth]{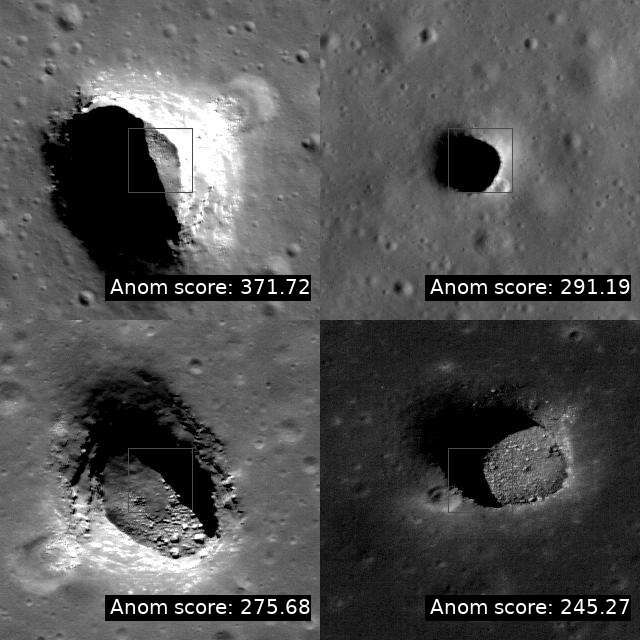}
    \caption{Volcanic Pits}
    \label{subfig:pits}
\end{subfigure}
\begin{subfigure}[t]{0.48\linewidth}
    \centering
    \includegraphics[width=\linewidth]{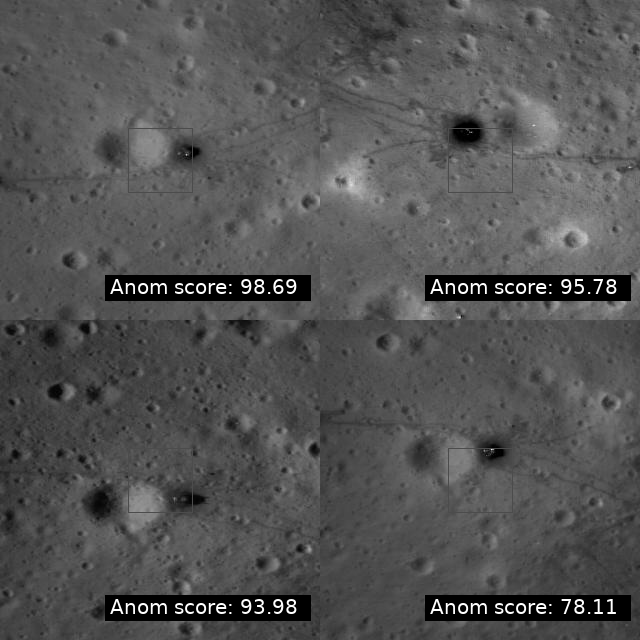}
    \caption{Apollo 16 lander}
    \label{subfig:apollo_16}
\end{subfigure} \hfill
\begin{subfigure}[t]{0.48\linewidth}
    \centering
    \includegraphics[width=\linewidth]{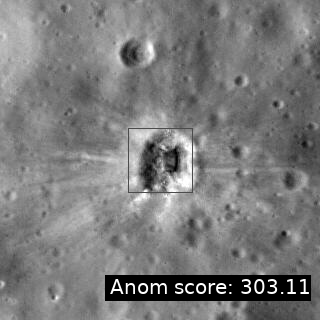}
    \caption{Apollo 13's S-IVB booster crash site}
    \label{subfig:ap_13_S_IVB}
\end{subfigure}
\begin{subfigure}[t]{0.48\linewidth}
    \centering
    \includegraphics[width=\linewidth]{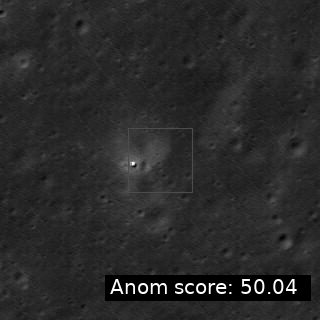}
    \caption{Chang'e 6 lander}
    \label{subfig:change_6}
\end{subfigure}
\caption{(a) The four most anomalous patches containing pits (collapsed lava tubes) | (b) The four most anomalous patches containing the Apollo 16 lander | (c) Crash site of Apollo 13's S-IVB booster rocket module | (d) One of the most recent lunar landers, China's Chang'e 6, on the far side of the Moon. Note that some images of the same anomaly can be flipped/rotated due to the spacecraft orientation at the time of imaging (top right image of (b); top left and bottom left images of (a))}
\end{figure}

Also, to expand upon the tests done in \cite{lesnikowski2024anomalies}, we fed an expanded suite of known landed technological assets into the model, including: Apollo $11$ lander, Apollo $12$ lander, Apollo $13$ S-IVB crash site, Apollo $14$ lander, Apollo $15$ lander, Apollo $16$ lander, Apollo $17$ lander, Surveyor $3$, Luna $24$ lander, SMART$1$ crash site, Chang'e $3$ lander, Change'e $5$ lander, and Chang'e $6$ lander. \autoref{fig:all_tech_anom_pr_and_kde} displays the results from the model examining images containing these technosignatures, again showing that despite their small size within the image patches, the model still identified the majority of patches containing technosignatures as anomalous when compared to the average lunar surface. While the anomaly scores of the artificial anomalies are not as high as natural anomalies (see \autoref{subfig:pits}), this can be partially explained by the size of the training patches, as the anomaly score calculation has a bias towards objects that dominate the patches.

\label{fig: anomaly_examples}

\section{Conclusion}

This investigation showed in a robust manner that the current version of the model can identify lunar surface anomalies well. Applying the model to a larger, if not global, dataset would unveil a plethora of anomalies. One of our team's next goals is to create a global surface anomaly map of the Moon in hopes of not only discovering previously unknown structures that could prove useful or interesting to future manned missions (volcanic pits, rockfalls, recent impact craters, etc.), but also missing or unknown landed assets. Newly identified anomalous patches could then even be used in a self-supervised manner to improve the model's performance.

This version of the model was trained based on certain hyperparameters that bias its performance, none more so than the patch size. By training more models on different sized patches, anomalies of different scales can be more readily identified. Combining these models would allow a more comprehensive examination of the lunar surface. The regularization term ($\beta$) could also be altered to make the recreated patch images less or more abstract. Testing different values of $\beta$ across different patch sizes would be another way to test performance across different biases.

Beyond completing a global lunar anomaly search, this technique can be applied to any Solar System body that has high-resolution imaging of its surface, such as Mars or, in the near future, Mercury. The model can even be utilized as a launching point to train new models on new data of a different object using a technique called transfer learning, where a learned model in a certain domain is used as a starting point to train a new model in a similar domain. The code is available on GitHub \footnote{\url{https://github.com/lesnikow/jstars-automated-discovery}} for the use of the community.

\bibliography{Sample.bib}

\end{document}